\documentclass[prl,twocolumn]{revtex4}
 \usepackage{graphicx, epsfig} \usepackage{color}
 \usepackage{mathrsfs}

 \newcommand{\be}{\begin{equation}} \newcommand{\ee}{\end{equation}}
 \newcommand{\bea}{\begin{eqnarray}}
 \newcommand{\eea}{\end{eqnarray}} 
 \newcommand{\gapp}{\mathrel{\raise.3ex\hbox{$$}\mkern-14mu
               \lower0.6ex\hbox{$\sim$}}}
 \newcommand{\lapp}{\mathrel{\raise.3ex\hbox{$<$}\mkern-14mu
               \lower0.6ex\hbox{$\sim$}}}

\begin{document}

 \title{Dark energy, the colored anti-de Sitter vacuum, and LHC phenomenology}

 \author{Dejan Stojkovic$^1$} \author{Glenn D. Starkman$^2$} \author{Reijiro
 Matsuo$^2$} \affiliation{$^1$HEPCOS, Department of Physics, SUNY at Buffalo, Buffalo, NY~~14260-1500\\ $^2$CERCA, Department
  of Physics, Case Western Reserve University, Cleveland, OH~~44106-7079}

 \begin{abstract} \noindent We study the possibility that the current
 accelerated expansion of the universe is driven by the vacuum energy
 density of a colored scalar field which is responsible for a phase
 transition in which the gauge $SU(3)_c$ symmetry breaks. We show that if we  are stuck in a $SU(3)_c$-preserving false vacuum, then $SU(3)_c$ symmetry breaking can be accommodated without violating any experimental QCD bounds or bounds from cosmological observations. Moreover, unless there is an unnatural fine-tuning beyond the
 usual cosmological constant fine-tuning, the true vacuum state of
 the universe is anti-de Sitter.  The model can likely be tested at the LHC.  A possible (though not necessary) consequence
 of the model is the existence of fractionally charged massive hadrons. The model can be embedded in supersymmetric
 theories where massive colored scalar fields appear naturally.
 \end{abstract}

  \maketitle

 Observational data indicate that our universe is going
 through a phase of accelerated expansion. To date it remains a
 mystery what is the driving force behind the acceleration. Data
 favor an equation of state of the cosmic fluid $w \approx -1$,
 corresponding to a constant, or nearly constant, energy density. The
 null hypothesis is that we have reached the lowest energy state of
 the universe -- the true vacuum energy density (or cosmological
 constant).

 If it is indeed the true vacuum energy we are seeing, it may
 represent the worst prediction ever made by a theory. The value
 needed to explain the observed acceleration, $(10^{-3}eV)^4$, is
 $124$ orders-of-magnitude smaller than the value naively predicted by particle theory, $(10^{19}GeV)^4$, and still $60$-ish
 orders-of-magnitude after the assistance of supersymmetry.

 It is equally plausible that the we are instead noticing the effects of
 some metastable false vacuum energy associated with some matter field.
 This would also have the equation of state $w=-1$. If this is the case,
 how might the true vacuum of the universe differ from the false vacuum
 in which we find ourselves?  One startling realization is that just because the
 magnitude of the current vacuum energy is so close to zero, does not
 mean that the magnitude of the true vacuum energy is similarly fine-tuned.
 Indeed, it would in many ways be more natural for the true
 vacuum energy to be separated from ours by an amount
 characteristic of whatever phase transition is associated with the
 transition from true to false vacuum. The true vacuum state of the universe
 might therefore be strongly anti-de Sitter.

 Non-zero vacuum energy density is usually
 associated with the breaking of some symmetry.
 What symmetry might be breaking?
 The most interesting possibility is that it is
 the unbroken gauge symmetry -- $SU(3))_c \times U(1)_{\rm EM}$.
 The Universe in its history has apparently undergone a number of
 phase transitions. Currently, the physics is
 invariant under local $SU(3)_c \times U(1)_{\rm EM}$
 transformations. However, there is no reason to believe that the
 chain of symmetry breaking stops here. We focus here on the
 possibility of $SU(3)_c$ breaking, though similar possibilities may
 exist for $U(1)_{EM}$ breaking.

 The possible breaking of $SU(3)_c$ symmetry was intensively
 discussed some time ago for different reasons
 \cite{color_breaking_old}. Depending on the model, the $SU(3)_c$
 gauge symmetry can be broken completely or down to
 some subgroup of $SU(3)_c$. Regardless of the details, a generic
 feature of $SU(3)_c$ breaking is the existence of a new
 colored scalar Higgs multiplet, $\Phi$. The non-zero vacuum
 expectation value of this field breaks the color symmetry and gives
 mass to gluons. To enforce the observed color symmetry locally, we
 will insist that the phase transition is first order, and that the
 observed universe is in a color-symmetric false vacuum.

 Consider a Lagrangian density invariant under the $SU(3)_c$
 transformations:
 \be
 \label{model} {\cal L} = -\frac{1}{2}{\rm Tr}
 F_{\mu \nu}F^{\mu \nu} + \frac{1}{2}{\rm Tr} (D_\mu \Phi) (D^\mu
 \Phi) -V(\Phi) \, .
 \ee
 For illustrative purposes, we take
 $\Phi$ in the adjoint representation of $SU(3)$.
 $D^\mu\Phi$ is the covariant derivative of
 $\Phi$ with a gauge coupling constant $g$.
 The potential
 \be
 V(\Phi) = \frac{\mu^2}{4}({\rm Tr}
 \Phi^2)+\frac{\lambda_1}{16}({\rm Tr}
 \Phi^2)^2+\frac{\lambda_2}{6}({\rm Tr} \Phi^3) + V_0\, .
 \ee
 A Hamiltonian bounded below is assured if $\lambda_{1}>0$.
 The term ${\rm Tr} \Phi^4$ is not included,
 but, since it has the same structure as $({\rm Tr} \Phi^2)^2$,
 it will not change our qualitative results.
 A similar model was studied in \cite{Viswanathan:1978yy}.
 We include a constant $V_0$, which is an overall
 shift of the potential.

 We choose to work in a diagonal representation where $\Phi$
 is Hermitian and traceless. Let the three real fields of the
 diagonal representation be $\psi_{1}$, $\psi_{2}$ and $\psi_{3}=-(\psi_{1}+\psi_{2})$.
 Minimizing V($\Phi$) we find
 \be
 \psi_{1}=\psi_{2}=-{\psi_{3}}/{2} \equiv \psi \, .
 \ee
 The potential as a function of $\psi$ is:
 \be \label{notuned}
 V(\psi)=\frac{3}{2}\mu^{2}\psi^{2}+\frac{9}{4}\lambda_{1}\psi^{4}-\lambda_{2}\psi^{3}
 +V_0 \, .
 \ee
 Defining
 \be \label{def}
 \psi_{0}\equiv\frac{2}{9}\frac{\lambda_{2}}{\lambda_{1}} \ \ \ \
 {\rm and } \ \ \ \
 \epsilon_{0}\equiv \lambda_{1} - \frac{2\mu^2}{3\psi_0^2} \, ,
 \ee
 allows us to rewrite $V(\psi)$ in a more convenient form:
 \be \label{Vepsilon}
 V(\psi)=
 \frac{9}{4}\lambda_{1}\psi^{2}(\psi-\psi_{0})^{2}
 -\frac{9}{4}\epsilon_{0}\psi_{0}^{2}\psi^{2}+V_0 \, .
 \ee

 The role of the parameter $\epsilon_0$ is to introduce a
 controlled fine tuning in $V(\psi)$.
 If $\epsilon_0=0$, $V(\psi)$
 has two degenerate minima at $\psi=0, \psi_{0}$.
 However, for a small positive $\epsilon_0$,
 the global minimum is at $\psi =\psi_0(1+\epsilon_0/\lambda_1)$.
 The difference between the energy densities of the two vacua is
 \be \label{eqn:deltaV}
 \delta V = 9\epsilon_0 \psi_0^4/{4} \, .
 \ee

 \begin{figure}[t]
 \includegraphics[width=0.65\linewidth]{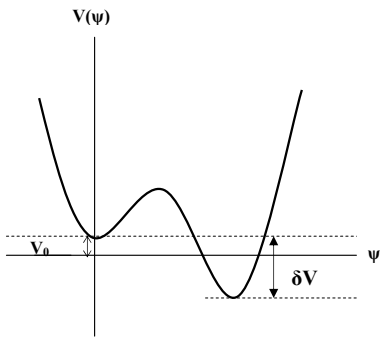}\\
 \caption{Characteristic potential for a first order phase
 transition. There exist the false vacuum where the symmetry is
 unbroken, $\psi =0$, and the true vacuum where the symmetry is
 broken, $\psi \neq 0$. $\delta V$ is the difference in
 energy densities between the vacua. $V_0$ plays the
 role of the cosmological constant.} \label{Vphi} \end{figure}

 We included $V_0$ in the potential so that we could specify the false
 vacuum energy density, $V(\psi=0)=V_0$. We require that $V_0$ is
 positive and extremely small, $V_0 \simeq (10^{-3}{\rm eV})^4$,
 representing the vacuum energy density that is driving the
 accelerated expansion of the universe. This situation is depicted in Fig. \ref{Vphi}.
 We do not explain the value of $V_0$ --
 it is an observational input,
 and its small value
 is the usual cosmological constant fine tuning,
 We will show, however, that that $\epsilon_0$ does
 not have to be fine-tuned:
  $\epsilon_0 \sim 0.01-0.1$ will be sufficient to avoid all constraints.
 Thus, the shape of the potential is not fine-tuned,
 Since generically $\delta V \gg V_0$, the true vacuum energy density will
 be large and negative, representing an AdS space.

 $V(\psi)$ is the zero-temperature potential.
 Since we are interested in the first-order phase transition between
 the false vacuum $\psi=0$, and the true vacuum $\psi=\psi_0$,
 we need the finite-temperature effective potential. Following
 \cite{Viswanathan:1978yy}, we find that at one-loop
 \be
 V_{\rm eff}(\psi)=
 \frac{9}{4}\lambda_{1}\psi^{2}(\psi-\psi_{0})^{2}
 -\frac{9}{4}\epsilon(T)\psi_{0}^{2}\psi^{2}+V_0 \, ,
 \ee
 where
 \be
 \epsilon(T) \equiv \epsilon_0 - \frac{2}{3}\frac{\mu^2(T) - \mu^2}{\psi_0^2}
 = \epsilon_0 - \left(\frac{5}{9}\lambda_1 + \frac{1}{2}g^2\right)\frac{T^2}{\psi_0^2}
 \, .
 \ee
 All temperature dependence is in
 $\mu^2(T)\equiv\mu^2 +(\frac{5}{6} \lambda_1 +\frac{3}{4}g^2)T^2$.
 The critical temperature at which $\epsilon(T)=0$ is
 \be
 T_{c}^2=\frac{2\epsilon_{0}\psi_{0}^2}{10\lambda_{1}/9+g^{2}} \, .
 \ee
 For $T\gg T_c$, the potential has a single minimum at $\psi =0$,
 color symmetry is thus unbroken. As $T$ falls,
 a second minimum emerges at $\psi \neq 0$;
 this becomes the global minimum at $T < T_c$.
 Thus, the true vacuum state breaks $SU(3)_c$; however,
 our universe is stuck in a supercooled color-preserving false vacuum
 with $\psi =0$. Thermal fluctuations at $T\ll T_c$ are not large enough
 to take the system over the barrier into the true vacuum.

 The existence of a lower minimum indicates that any point
 in the universe will eventually tunnel into the true $SU(3)_c$-breaking vacuum.
 However, the tunneling rate depends on the energy difference between the
 vacua and the height of the potential barrier, and can be extremely
 slow as we discuss below. A universe in
 such a false vacuum will evolve as has ours -- initially radiation
 dominated, then matter dominated, and eventually expanding at an
 accelerated rate as it becomes dominated by the
 false vacuum energy density, $V_0$.

 In the false vacuum state, $\psi=0$, excitations of the colored Higgs
 have mass $m_\psi \sim \mu$. It is natural to chose $\mu$ to be
 ${\cal O}(1{\rm TeV})$ as this is the energy scale
 at which new physics beyond the Standard Model is expected.
 Without new physics at this scale, the Standard Model is severely fine-tuned.
 It is also very difficult to hide a colored scalar field much lighter that this.
 This choice automatically evades all of the experimental QCD
 constraints.

 The energy density of the false vacuum must, as we noted, be
 chosen phenomenologically,
 $V_0 \equiv \rho_v \approx (10^{-3} {\rm eV})^4$.
 However, an interesting numerology,
 $V_0 \approx \left( {\rm TeV}^2/M_{{\rm Pl}}\right)^4$,
 hints perhaps toward a gravitational origin of the
 overall shift of the potential if the fundamental scale of the
 physics is TeV.

 We next examine the limits on the parameter $\epsilon_0$. An
 important question is how long would the universe exist in the
 supercooled false vacuum state $\psi=0$.
 The transition from this de Sitter-like false
 vacuum to the true anti-de Sitter-like vacuum, is realized by
 nucleation of bubbles of true vacuum inside the false vacuum.
 In the semi-classical approximation,
 the transition probability per unit space-time volume is
 \be \label{prob}
 \Gamma = B e^{-S_E}
 \ee
 where $S_{\rm E}$ is the Euclidean action of the
 $O(4)$-symmetric bounce solution describing the tunneling.
 $B$ is a dimensionful constant ${\cal O}({\rm TeV}^4)$, which depends on quantum loop corrections to the potential (\ref{Vepsilon}). Since we are interested only in the order of magnitude of the transition rate, we ignore these corrections. To calculate $S_{\rm E}$,
 we follow the method developed in \cite{Coleman:1977py}.

 The one-dimensional, zeroth-order (in $\epsilon_0$) Euclidean action per
 unit volume for the tunneling is
 \be
 S_1 = \int_{0}^{\psi_0}
 d\psi' \sqrt{2V (\psi')} \approx \sqrt{\frac{\lambda_1}{8}} \psi_0^3\,
 .
 \ee
 In the zero-temperature limit and thin wall approximation, the
 radius of the critical bubble is
 \be
 \label{critical_bubble} R_{0}=\frac{3S_{1}}{\delta V} =
 \frac{\sqrt{2\lambda_{1}}}{3}\frac{1}{\epsilon_0\psi_{0}} \, .
 \ee
 The Euclidean action for an $O(4)$ symmetric bubble is
 \be
 S_{E}=-\frac{1}{2}\delta V\pi^{2}R_{0}^{4}+2\pi^{2}R_{0}^3S_{1}
      =\frac{\pi^{2}\lambda_{1}^{2}}{54}\frac{1}{\epsilon_{0}^{3}}
 \ee
 At zero temperature, the decay rate per unit volume per unit
 time is given by (\ref{prob}). In order for our observable
 universe, whose four-volume is of the order of $t_{\rm Hubble}^4$,
 to remain in the false vacuum, one needs roughly $\Gamma t_{\rm
 Hubble}^4 \lapp 1$. Taking  $t_{\rm Hubble} \sim 10^{10}$years, we find that
 sufficient meta-stability of the color-neutral false vacuum is
 obtained for $S_{\rm E} > 400$. With a generic value
 $\lambda_1 \sim 1$, we see that vacuum stability
 requires only $\epsilon \lapp 0.1$, an extremely mild fine-tuning.
 On the other hand, $\epsilon \lapp 0.1$ is enough to make the thin
 wall approximation valid.

 To make sure that the story remains unchanged in the early universe,
 we calculate the temperature dependent decay rate in the high temperature limit.
 At finite temperature, instead of an $O(4)$ symmetric bounce,
 one should look for an $O(3)$ symmetric solution,
 periodic in time with period $1/T$ (see \cite{Linde}).
 In the high temperature limit ($T \gg 1/R_{0}$),
 the time integration in the calculation of the bounce action
 yields an overall factor of $1/T$.  The decay exponent $S_E$ now has the form
 \be
 S_E={S_{3}(\psi,T)}/{T} \, ,
 \ee
 where
 $S_{3}$
 is the three-dimensional action of the $O(3)$ symmetric bubble.
 The radius of the critical bubble is
 \be
 R(T)=\frac{2S_{1}}{\delta V(T)}
     =\frac{\sqrt{8\lambda_{1}}}{9}\frac{1}{\epsilon(T)\psi_{0}} \, .
 \ee
 The three dimensional action of the bounce is given by
 \begin{eqnarray}
 S_{3}(T)&=&-\frac{4}{3}\pi R(T)^{3}\delta V(T)+4\pi R(T)^{2}S_{1}\\
 &=&\frac{\sqrt{128\lambda_{1}^{3}}}{243}\frac{\psi_{0}}{\epsilon^{2}(T)}\nonumber
 \end{eqnarray}
 The temperature dependent decay rate is now
 \be \label{tddc}
 \Gamma(T) \approx
 \exp\left[-\frac{\sqrt{128\lambda_1^{3}}}{243}\frac{\psi_{0}}{T\epsilon^{2}(T)}\right]
 \ee

 Thus, at high temperatures $T \gg T_c$, when the color preserving
 minimum at $\psi =0$ is the lowest energy state, the transition rate
 is large and most of the universe ends up in that minimum.
 At $T= T_c$, we have $\epsilon(T)=0$ and transitions between the vacua are suppressed. The high temperature approximation is
 valid, at least formally, at the temperatures slightly below $T_c$.
 It is there that the decay rate (\ref{tddc}) is maximal and we need to correct the zero temperature estimate for $\epsilon_0
 $. Fortunately, a slight correction $\epsilon_0 \sim 0.05$ makes the decay rate safely small.
 As shown earlier, for $T \ll T_c$, $\psi =0$ is a false minimum, but
 the transition rate to the true color-breaking vacuum is suppressed by the bare value of $\epsilon_0$. We saw that $\epsilon
 _0 \sim 0.1$
 makes the transition time larger than the current Hubble time. In order to incorporate somewhat stronger constraint for high
  temperatures, we require $\epsilon_0 \sim 0.05$.

 One might worry that, even at $T \ll T_c$, energetic processes
 in the history of our universe (e.g. cosmic ray collisions) would
 have stimulated the formation of a true vacuum bubble, and that such
 a bubble would have expanded rapidly to encompass most of the
 visible universe. Fortunately, it is not a simple thing to create a
 vacuum bubble in a high energy collision. This requires not
 just sufficient energy but a coherent superposition of a large
 number of high energy quanta over a volume large compared to the
 characteristic energy. Such processes happen freely at high
 temperature but essentially not at all at high energy.
 The height of the barrier between the false and true vacua is
 \be \label{eqn:Vmax}
 V_{max} =
 {9}\lambda_{1}\psi_{0}^{4}/{64} \, .
 \ee
 This is approximately
 TeV$^4$ for $\psi_0\sim 1$TeV and $\lambda_1 \sim 1$. Such
 temperatures are unlikely to soon be achieved in colliders,
 and are probably not achieved over large enough volume
 even in the highest energy cosmic ray collisions.

 The critical bubble radius (\ref{critical_bubble})
 is $5{\rm TeV}^{-1}$ for our canonical parameters
 ($\epsilon_0 \sim 0.1$, $\psi_0\sim 1$TeV and $\lambda_1 \sim 1$).
 Given (\ref{eqn:Vmax}), this suggests that we need approximately
 \be
 N_{quanta} \simeq
 \left( \frac{4\pi R_0^3}{3} V_{max} \right) {V_{max}^{-1/4}}  > 100 \left({0.1}/{\epsilon}\right)^3.
 \ee
 individual excitations coherently supposed.  Note that $N_{quanta}$ grow
 very fast -- as $\epsilon^{-3}$.
 To create a bubble in a high energy collision is thus extremely difficult.
 A bubble is a highly coherent state of $\Phi$ particles.
 Produce these quanta and they are more likely to fly apart
 than to assemble into a true vacuum bubble.
 The probability that you make a bubble should therefore be suppressed
 by a huge entropy factor.
 Cosmic rays,
 which usually produce collisions with $E_{cm}\lapp 100$TeV,
 are thus extremely unlikely to trigger the destruction of the known universe.

 The  scenario proposed here should have distinct experimental signatures
 in near-future accelerators like the Large Hadron Collider (LHC) and
 in cosmic ray observatories.
 The principal generic feature of our scenario is
 the existence of colored scalar fields with mass around $1$TeV.
 Experimental signatures depend crucially on the lifetimes of
 excitations of these fields, which are model dependent.
 A heavy colored scalar can generically decay into gluons and quarks.
 Its lifetime depends on couplings and on the specific representation
 of $SU(3)_c$. If the decay is fast, the main signature of its production will be
 gluon and quark jets in excess of the standard model prediction.

 The lifetime of the colored scalar could be longer than the characteristic
 hadronization time $\sim (100 {\rm MeV})^{-1}$.
 This could happen if $\Phi$ is in a high-dimensional representation of $SU(3)_c$
 or is protected by a symmetry, such as R-parity.
 In an unbroken phase, it is impossible to find an isolated colored particle;
 however, color singlets can be found free,  thus $\Phi$ would combine into
 color singlet bound states.
 One possibility is bound state of two or more colored scalars
 (depending on the $SU(3)_c$ representation to which $\Phi$ belongs).
 It was argued in \cite{Dawson:1982cp} that such states would resemble
 glueballs (bound states of gluons).
 Another accessible new state is a bound state of $\Phi$ with quarks.
 If the scalar particle does not carry electromagnetic charge,
 these states could be fractionally charged --
 a unique (though not generic) signature of our model.
 We would expect the lightest of these new states
 (e.g. a bound state of $\Phi$ with the lightest quarks)
 to have mass of the order of a few TeV.

 Specific experimental signatures will depend on the exact model of $SU(3)_c$ breaking.
 For example, in the model discussed above,
 the symmetry breaking is realized by a scalar field that transforms
 under the eight-dimensional adjoint representation.
 Since $8 \times 8$ in the $SU(3)_c$ representation classification scheme
 contains an $SU(3)_c$ singlet state, one can find a color-free bound
 state of two scalar fields. Like glueballs, such states
 would dominantly decay into lighter mesons. Quarks (q) and
 antiquarks ($\bar{q}$) belong to $3$ and $\bar{3}$ representations of
 $SU(3)_c$. The simplest combination that contains a color singlet is
 $8 \times 3 \times \bar{3}$, corresponding to $\Phi q
 \bar{q}$. This combination has integer charge.
 However, in models where $SU(3)_c$ is broken by a scalar multiplet
 belonging to a $3, 6, 10, 15$ or $21$ dimensional representation,
 the bound states can have fractional charges.
 The $27$ of $SU(3)$ again gives an electrically neutral bound state.
 Integer charged states decay to lighter mesons,
 whereas the lightest fractionally charged state is presumably stable.
 These experimental signatures will be explored in greater detail in future publications.

 While discovery of a colored scalar field would be indicative,
 it would shed no light on the global shape of the potential.
 In \cite{Vachaspati:2003vk}, a method for reconstructing the potential was proposed.
 In a model containing kink solutions,
 one could in principle characterize the potential by studying the spectrum of kinks.
 Producing a kink corresponds to producing a small bubble of true vacuum.
 Since the potential barrier is only a TeV high, it might be possible to realize
 this in future experiments.
 Such bubbles would be subcritical, only TeV$^{-1}$ in size,
 and would decay quickly. The decay pattern of subcritical bubbles
 would shed light on the global shape of the potential.
 Production of a large, supercritical bubble is suppressed.
 Thus, it might be possible to test the model without destroying the known universe.

 Finally we mention the possibility that an $SU(3)_c$-breaking
 true vacuum can be embedded into supersymmetric models.
 Supersymmetric models naturally contain
 colored massive scalar fields as super-partners of the ordinary
 standard model quarks. It has been thought that such fields
 are dangerous since they can lead to breaking of the $SU(3)_c$
 symmetry, {\it i.e.} the true minimum in the landscape of supersymmetric minima
 may be one with broken $SU(3)_c$.
 A large portion of the supersymmetric parameter space
 is therefore excluded \cite{Casas:1995pd}.
 Based on the discussion presented here, we see that one does not
 need to do this. In particular, false minima where $SU(3)_c$ is not broken and
 which lie above the true vacuum can drive the accelerated expansion
 of our universe (for related work see \cite{Kusenko:1996jn,Abel:1998ie}). Note
 that in this case $\Phi$ is likely a squark field and so
 carries fractional electric charge. $\Phi$ cannot however
 be the dark matter (nor can the dark matter be a bound state involving
 $\Phi$, as that would contradict well-known limits on strongly
 interacting dark matter \cite{Starkman:1990nj}).

 In conclusion, we have proposed that our universe is stuck in a false vacuum
 where $SU(3)_c$ symmetry remains unbroken, while the true vacuum breaks $SU(3)_c$.
 The vacuum energy density of the false vacuum drives the observed accelerated expansion.
 A slight tuning of the potential, $\epsilon_0 \sim 0.01-0.1$,
 makes the transition time to the true vacuum larger than the Hubble time and provides a self-consistent high temperature history of our universe.
 Massive colored scalar fields provide unique experimental
 signatures in near-future accelerator experiments.
 Discovery of fractionally charged hadrons would be particularly strong evidence for our model.
 Using recently developed techniques, one could in principle determine the
 shape of the potential by solving the inverse scattering problem.
 The model can be embedded in supersymmetric
 theories where massive colored scalar fields appear naturally.
 We also note that with small modifications,
 a similar mechanism can be applied to $U(1)_{\rm EM}$, the
 gauge group of electromagnetism or to the conventional Higgs of the Standard Model.

 \begin{acknowledgments} The authors are grateful to T.~Vachaspati,
 C.~Csaki and M.~Cvetic for useful discussions. This work was
 supported by the HEPCOS group at SUNY at Buffalo and the U.S. DoE at CWRU.
 \end{acknowledgments}

 \end{document}